\documentclass[aps,pra,altaffilletter,superscriptaddress,twocolumn,showpacs]{revtex4}

\usepackage{amsmath}
\usepackage{amssymb}
\usepackage{graphicx}
\usepackage{color}
\usepackage{bm}

\newcommand{\dd}{\mathrm{d}}
\newcommand{\ee}{\mathrm{e}}
\newcommand{\ii}{\mathrm{i}}

\newcommand{\calO}{\mathcal{O}}

\usepackage{dcolumn}                 % Aligning table columns
\newcolumntype{w}[1]{D{.}{.}{#1}}

\newcommand*{\cent}[1]{\multicolumn{1}{c}{$#1$}}

\newcommand{\addrMissouri}{Missouri University of Science and
Technology, Rolla, Missouri 65409-0640, USA}

\newcommand{\addrNIST}{National Institute of Standards and Technology,
Gaithersburg, Maryland 20899-8420, USA}

\newcommand{\addrPoznan}{Faculty of Chemistry,
Adam Mickiewicz University, Grunwaldzka 6, 60-780 Pozna\'n, Poland}

\begin{document}

%
% Quantum Electrodynamic Corrections to the g Factor of Helium P States
%
\title{Quantum Electrodynamic Corrections to the $\bm{g}$ 
Factor of Helium $\bm{P}$ States}

\author{M. Puchalski}
\email{mpuchals@fuw.edu.pl}
\affiliation{\addrPoznan}
\affiliation{\addrMissouri}

\author{U. D. Jentschura}
\email{ulj@mst.edu}
\affiliation{\addrMissouri}
\affiliation{\addrNIST}

\begin{abstract}
The Land\'{e} $g$ factor describes the response of an atomic energy level to an
external perturbation by a uniform and constant magnetic field.  In the case of
many-electron systems, the leading term is given by the interaction $\mu_B \, ( \vec L
+ 2 \, \vec S) \cdot \vec B$, where $\vec L$ and 
$\vec S$ are the orbital and spin angular momentum 
operators, respectively, summed over all electrons. For helium,
a long-standing experimental-theoretical discrepancy 
for $P$ states motivates a reevaluation of the 
higher-order terms which follow from relativistic 
quantum theory and quantum electrodynamics (QED).
The tensor structure of
relativistic corrections involves scalar, vector, and symmetric and
anti-symmetric tensor components. We perform a tensorial reduction 
of these operators in a Cartesian basis, using an approach which allows 
us to separate the internal atomic from the external degrees of freedom
(magnetic field) right from the start of the calculation.
The evaluation proceeds in a Cartesian
basis of helium eigenstates, using a weighted sum over the 
magnetic projections.
For the relativistic corrections,
this leads to a verification of previous results obtained using the
Wigner--Eckhart theorem. The same method, applied to the radiative correction
(Bethe logarithm term) leads to a spin-dependent correction which is different
for singlet versus triplet $P$ states. Theoretical
predictions are given for singlet
and triplet $2P$ and triplet $3P$ states and compared to 
experimental results where available. 
\end{abstract}


\maketitle

%
% Introduction
%
\section{Introduction and Overview}

%
% Basics of the $g$ factor in few-electron systems
%
\subsection{Few-electron systems and $\bm{g}$ factor}

The quantum electrodynamic (QED) theory of bound systems describes, among other
things, three ``fundamental'' characteristic effects that involve the spectrum
of bound systems, namely, (i)~the Lamb shift, which is the energy shift of
bound states due to the self-interaction of the electrons, and due to tiny
corrections to the Coulomb force law at small distances, (ii)~the $g$
factor of bound states, which describes the energy shift of a bound state due
to the interaction with an external, uniform magnetic field (Zeeman effect),
and (iii)~the hyperfine splitting, which is given by the interaction of 
bound electrons with the nuclear magnetic moment.  These effects seem to be the
three most commonly studied QED effects for bound states, because of prominent
high-precision experiments in all three mentioned areas.  The leading QED
corrections to all three mentioned effects are given by the self-energy 
of the orbiting particle, and by vacuum polarization. 

The long standing discrepancy between
theory and experiment for the Zeeman coupling factor $g'_L$ for
the $2^3P$ state of helium~\cite{LhFaBi1976} has motivated a number of 
independent theoretical papers~\cite{LeHu1975,AnSe1993,YaDr1994,Pa2004}
on the subject. Here, by convention, $g'_L$ is the complete 
orbital part of the $g_J$ factor for the helium $P$ state,
including relativistic and radiative corrections.
For hydrogenlike systems, the self-energy 
corrections to the $g_J$ factor and to the hyperfine splitting 
can be formulated in a similar framework~\cite{YeJe2008,YeJe2010},
by observing that they can be
described as a ``dressed'' self-energy correction in an additional 
magnetic field, namely, for the case of the $g_J$ factor, 
in a uniform external magnetic field, and, 
for the case of the hyperfine splitting, 
in the magnetic dipole field of the atomic nucleus.

For more complex atoms and ions, the theory of the $g$ factor is more
complicated because in higher order, the electron-electron interaction
is intertwined with the coupling to the external magnetic field.
In leading order, the total orbital angular momentum $\vec L$ 
and the spin angular momentum $\vec S$ couple to the external 
magnetic field $\vec B$ as described by the Hamiltonian matrix element
\begin{equation}
\left< H_M \right> \approx
\left< \mu_B \, (\vec L + 2 \, \vec S) \cdot \vec B \right> = 
g_J \, \mu_B \, B \, \mu \,,
\end{equation}
where $g_J$ is the Land\'{e} $g$ factor, and 
$\mu_B$ is the Bohr magneton, that is,
$\mu_B = -e/(2 m)$ where $m$ is the electron mass and 
$e = -|e|$ is the electron charge.
The orbital angular momentum $\vec L$ and the 
spin angular momentum $\vec S$ are summed
over all electrons. As long as the 
separation into terms proportional to 
$\vec L \cdot \vec B$ and
$\vec S \cdot \vec B$ remains valid,
this gives rise to an orbital $g_L \approx 1$ factor
and a spin $g_S \approx 2$ factor, so that 
\begin{align}
\label{conv}
g_J =& \; g_L \,
\frac{J (J+1) + L (L + 1) - S(S+1)}{2 J(J+1)} 
\nonumber\\[0.77ex]
& \; + g_S \,
\frac{J (J+1) + S (S+1) - L (L + 1)}{2 J(J+1)}  \,.
\end{align}
In leading order, the Land\'{e} $g$ factor is thus given by the well-known
formula
\begin{align}
g_J \approx & \; \frac{3 J (J+1) - L (L + 1) + S(S+1)}{2 J(J+1)} \,.
\end{align}
In higher order, due to spin-orbit coupling, one cannot separate 
the magnetic-field interaction any more into terms 
proportional to $\vec L \cdot \vec B$ and
$\vec S \cdot \vec B$, and therefore, one cannot 
uniquely identify the orbital $g_L$ and spin $g_S$ factors 
any more. For hydrogen, the corresponding mechanism
has been discussed in Appendix A of Ref.~\cite{Je2010g}. 
The separation into $g_L$ and $g_S$ remains valid 
up to relative order $\alpha^3$, where $\alpha$ is the 
fine-structure constant,
provided one adds a tiny correction due to a
higher-order tensor structure, called $g_x$
in Refs.~\cite{LePiHu1970,YaDr1994}.

%
% Angular-momentum algebra
%
\subsection{Angular-momentum algebra}

For $P$ states as opposed to $S$ states, the angular momentum algebra involved
in the calculation of the bound-electron $g$ factor can become rather
complicated, and two approaches have been used. In approach (i), used in
Refs.~\cite{LePiHu1970,YaDr1994}, the authors formulate the entire theory in
terms of Wigner $3J$, $6J$, and $9J$ symbols, which enables them to perform all
calculations in terms of reduced matrix elements.  In turn, these can be
written in terms of the radial component of the wave functions as obtained from
variational calculations.  

In approach (ii), which has been used for hydrogenlike systems~\cite{Je2010g},
one first chooses a specific component of the Hamiltonian ``vector'' $\mu_B
(\vec L + 2 \vec S)$ multiplying the magnetic field $\vec B$, and a specific
magnetic projection of the reference state. Natural choices consist in the $z$
component of the Hamiltonian ``vector'' and the state with magnetic projection
$\mu = \tfrac12$ as indicated in Eqs.~(15) and~(16) of Ref.~\cite{Je2010g}.
Due to the Wigner--Eckhart theorem, one can then formulate all relative
corrections to the $g$ factor in terms of ratios, relating the effect
calculated with a correction to the magnetic Hamiltonian to the leading-order
effect, provided one uses the same state for each matrix element.  This
disentangles the internal atomic degrees of freedom from the external degrees
of freedom (the magnetic field).  For the hyperfine splitting, a similar
approach is outlined around Eq.~(7) of Ref.~\cite{YeJe2010}.

For helium, it is preferable to formulate the theory in terms of elements of
the radial wave functions alone, by expressing the matrix elements in terms of
sums over magnetic projections, where the angular and spin degrees of freedom
are summed over and evaluated in closed form. The latter sum can naturally be
expressed in terms of a ``radial'' representation of a $P$ state as obtained
from a variational calculation in a fully correlated, nonrelativistic basis.
Here, we thus choose an approach combining ideas from (i) and~(ii). First, the
relativistic and radiative corrections are expressed in terms of particular
tensor structures, and then, we evaluate these on a weighted sum over the
projections $m$ of the total angular momentum of the helium state.  This
approach combines the advantages of approach~(i), namely, the easy
applicability to helium, with the advantages of approach~(ii), namely, the full
disentanglement of the external degrees of freedom (magnetic field) from the
internal atomic degrees of freedom right from the start of the calculation.

Our investigation is motivated in part by an interesting
theoretical-experimental disagreement between the experimental result 
reported in Ref.~\cite{LhFaBi1976} and theory work 
described in Refs.~\cite{LeHu1975,AnSe1993,YaDr1994} and 
Sec.~V of Ref.~\cite{Pa2004}.
Our calculation is valid up to and including relativistic 
and radiative correction of relative order $\alpha^3$, and to second order
in the electron-nucleus mass ratio (for the leading nonrelativistic term). 
We proceed as follows. In Sec.~\ref{ham}, the
terms in the Hamiltonian which govern the bound-state $g$ factor are analyzed
in terms of their tensor structure. The discussion is complemented in
Sec.~\ref{GJ} by an analysis of the spin and the tensor reduction of the
particular correction terms.  Finally, in Sec.~\ref{num}, numerical evaluations
are described which allow us to obtain a highly accurate theoretical
prediction for the $g_J$ factor in helium, for $2P$ and $3P$ states.
Conclusions are reserved for Sec.~\ref{conclu}. Atomic units
with $e = \hbar = 1$ ($e$ denotes the physical electron charge), 
unit electron mass $m = 1$,
$\alpha = 1/c$ and $\epsilon_0 = 1/(4 \pi)$ are used throughout the paper.

%
% Hamiltonian
%
\section{Hamiltonian} \label{ham}

\subsection{Leading order}
\label{leading_order}

A careful treatment of the $g$ factor requires an analysis of the 
reduced-mass dependence. We denote the electron mass as $m$
and the mass of the nucleus as $M$.
The reduced mass $\mu$ and the mass ratio $\lambda$ are given by
\begin{equation}
\label{deflambda}
\mu = \frac{m M}{m + M} \,, 
\qquad
\lambda = -\frac{\mu}{M} \,,
\end{equation}
The interaction with the external magnetic field,
in leading order plus the 
reduced-mass correction, is given by~\cite{Ph1949,Pa2008}
\begin{align}
\label{HM}
H_M =&\; \mu_B\,\sum_a \, 
\big[g_L\,(\vec r_a \times \vec p_a) + 
\frac{g_S}{2}\,\vec \sigma_a  \big] \cdot \vec B  \nonumber \\ 
&\;-\mu_B\,\frac{m}{M}\,\sum_{a\neq b} \,
(\vec r_a \times \vec p_b) \cdot \vec B \,.
\end{align}
The finite mass of the nucleus yields a correction term 
(second term) of order $\calO(\lambda)$.
The sum over $a$ and $b$ in Eq.~\eqref{HM} 
counts the electrons of the bound system.
The well-known spin factor $g_S$ can be
expressed in the form (including two-loop corrections)
\begin{equation}
g_S = 2\, \left(1 + \frac{\alpha}{2\pi} -
0.328\,478\,695\,\frac{\alpha^2}{\pi^2} + \ldots\right) \,.
\end{equation}
It is equal to the $g$ factor of the free electron 
including the anomalous magnetic moment. The terms proportional to 
$g_L$ in Eq.~\eqref{HM} give rise to an orbital factor 
$g_L = 1 -m/M$. The terms in Eq.~\eqref{HM} contain all terms 
of relative order $\calO(\lambda)$. 
We note that this scaling of $g_L$, which has originally been derived 
in Ref.~\cite{Ph1949}, goes beyond the `trivial scaling' 
of momenta and distances which is discussed below.

Namely, in general, the scaling of the momenta and distances
with the reduced mass entails the 
scaling factors (see Appendix~\ref{appa})
\begin{equation}
\vec p \rightarrow \vec p\,(1 + \lambda) \,,
\qquad
\vec r \rightarrow (1 + \lambda)^{-1}\, \vec r \,.
\end{equation}
It results in prefactors of the form $(1+\lambda)^n$
with a certain scaling degree $n$. 
For the leading terms given in Eq.~\eqref{HM}, we have $n=0$,
and the terms commute
with the nonrelativistic Hamiltonian of the helium atom 
[without mass polarization; see Eq.~\eqref{A2} below]
\begin{equation}
\label{H0}
H_0 = \sum_a \left( \frac{\vec p_a^2}{2\mu} - \frac{Z}{r_a} \right) 
+ \sum_{a > b} \frac{1}{r_{ab}} \,.
\end{equation}
While the first-order correction to the wave function due to the 
magnetic interaction vanishes, the mass polarization term $H_{\rm mp}$ 
\begin{equation}
\label{Hmp}
H_{\rm mp} = -\frac{\lambda}{\mu}\, \sum_{a>b} \vec p_a \cdot \vec p_b \,.
\end{equation}
generates a nonvanishing perturbation to the wave function.
The perturbation can then be evaluated on the 
leading-order Hamiltonian~\eqref{HM}.
For the finite mass effect of order $O(\lambda^2)$, 
one additional effect is the mass polarization correction to the second term in
Eq.~\eqref{HM}. The third-order term involving the 
leading magnetic interaction term in Eq.~\eqref{HM} and two mass polarization 
insertions also yields a finite-mass correction
of second order in~$\lambda$.
Its effect on the $g_L$ prefactor and off-diagonal 
corrections are discussed in the following.

%
% Tensor decomposition of the Zeeman Hamiltonian
%
\subsection{Tensor decomposition of the Zeeman Hamiltonian}
\label{tensor}

Let us now turn to the tensor decomposition of the
Zeeman Hamiltonian. 
The first term in Eq.~\eqref{HM} can be rewritten in the form
\begin{subequations}
\label{HM0}
\begin{align}
H_{M0} = &\; \mu_B\,\vec G_0 \cdot \vec B\,, \\
G^i_0 = &\; \sum_a \,\left( g_L\,v_{0,a}^i + 
\frac{g_S}{2}\,d_0\,\sigma_a^i\right) \,, \\
v_{0,a}^{i} = &\; (\vec r_a \times \vec p_a)^i\,, \qquad d_0 = 1 \,,
\end{align}
\end{subequations}
where $v_{0,a}^{i}$ is a vector coefficient 
and $d_0$ is a diagonal (scalar) coefficient
multiplied only by a spin matrix.
Here and in the following,
Cartesian coordinates are denoted by superscripts,
that is, the $x$ component of $\vec v_{0,a}$ is
given as $v^x_{0,a}$ (the superscript assumes the values 
$i = x,y,z$).
We have decomposed the tensor structure of Eq.~\eqref{HM0}
into a vector and a spin part.
This approach is now generalized 
to other corrections $\delta H$ to the leading Zeeman Hamiltonian $H_{M0}$,
\begin{equation}
\delta H_M = \mu_B\,\sum_\gamma \vec G_\gamma \cdot \vec B \,,
\label{deltaHM}
\end{equation}
where $\gamma$ counts the correction terms.
The operators $\vec G_\gamma$ are 
linearly coupled to the magnetic field $\vec B$. We
split each element $G_\gamma$ into a tensor structure of spatial coordinates
coupled to the magnetic field $\vec B$ as well as
spin matrices $\vec \sigma_a$.
From the spinless terms of the form $\vec v \cdot \vec B$,
we obtain the vector coordinates~$v^i$. 

The second-order spatial tensors $A^{ij}$ 
in terms of the form 
$A^{ij} \sigma_a^i B^j $ can be tensorially decomposed into
a diagonal (scalar) part $d$, 
a symmetric tensor part $t$, and 
an antisymmetric tensor part $r$, 
\begin{subequations}
\label{tensor2}
\begin{align}
A^{ij} =& \; \frac{d}{3} \, \delta^{ij} + t^{ij} + r^{ij} \,, \\
d =&\; A^{kk}\,, \\
t^{ij} =&\; \frac{A^{ij} + A^{ji}}{2} - \frac{1}{3}\delta^{ij}\,A^{kk}\,,\\
r^{ij} =&\; \frac{A^{ij} - A^{ji}}{2}\,,
\end{align}
\end{subequations}
where the summation convention is used for the 
Cartesian coordinates as is done throughout 
the paper. 

The orbital angular momentum part in leading order is identified as 
the vector term $v^i_{0,a}$, and the spin part as related
to the scalar operator $d_0$. 
For the finite mass correction in Eq.~\eqref{HM} with
$G_1^i$ in tensor form, we have the identification
\begin{subequations}
\label{G1}
\begin{align}
G_1^i =& \; -\frac{m}{M} \, \sum_{a \neq b} \,v_{1,ab}^{i}\,, \\
v_{1,ab}^{i} =&\; (\vec r_a \times \vec p_b)^i\,,
\end{align}
\end{subequations}
which is included as the first term ($\gamma=1$) in $\delta H_M$. 
This grouping is extended to higher-order
terms and to make contact with the literature. 
According to Appendix~A of Ref.~\cite{LePiHu1970} and 
Eqs.~(2)---(4) of Ref.~\cite{YaDr1994},
we can split the $g_L$ and $g_S$ factors into 
leading-order terms, denoted by the same symbols, and 
corrections $\delta g_L$ and $\delta g_S$, which, 
when added to $g_L$ and $g_S$, 
yield the complete results $g'_L$ and $g'_S$,
which include the correction terms. So, for triplet $P$ states,
\begin{equation}
\label{defdeltag}
\delta g_L = g'_L - g_L \,, \qquad
\delta g_S = g'_S - g_S \,.
\end{equation}
This notation has been
introduced in the theoretical analysis of the experimental data for helium
$2^3P$-states~\cite{LePiHu1970} based on angular momentum methods
\cite{YaDr1994,InUf1958,Ed1974ang}. 
Compared to Eqs.~(22) and~(23) of Ref.~\cite{YaDr1994}, the prefactors in the 
expressions $\delta g_L = g'_L - \sqrt{(2L+1)L(L+1)/6} \; g_L$ and
$\delta g_S = g'_S - \sqrt{(2S+1)S(S+1)/6} \; g_S$ evaluate to unity for 
triplet $P$ states; for singlet $P$ states, the first equality 
in Eq.~\eqref{defdeltag} remains valid while $g_S = 0$.
The symmetric tensor parts $t^{ij}$ are related  to the $g_x$
factor~\cite{LePiHu1970}, and the mean values of the antisymmetric part $r^{ij}$
result in a zero correction. Later, these quantities were determined in
subsequent theoretical calculations of other authors \cite{He1973,YaDr1994}.
We follow these conventions in order to be able to compare 
our final formulas with their results. 
We here use a Cartesian decomposition of the 
higher-order Zeeman Hamiltonian, 
as an alternative to angular algebra methods 
with $3J$, $6J$, and $9J$ symbols~\cite{LePiHu1970},
and identify the tensor contributions to $g_L$, $g_S$, $g_x$ 
as described in the following.

\begin{widetext}
%
% Relativistic Corrections
%
\subsection{Relativistic corrections}
\label{rel_corr}

Relativistic corrections have been derived from the Breit Hamiltonian
\cite{PeHu1953,AbvV1953,KavV1954}.  We follow formulas from Eq.~(32) of
Ref.~\cite{Pa2004} with six relativistic corrections to the Zeeman effect,
\begin{align}
\label{Hrel}
& \delta H_{\rm rel} =
\mu_B\,\alpha^2 \, \sum_a
\bigg\{-\frac{\vec p_a^{\,2}}{2} \big[(\vec r_a \times \vec p_a) 
+ \vec \sigma_a \big]\cdot \vec B 
+ Z\,\frac{g_S - 1}{4}\,
\frac{(\vec r_a \times \vec \sigma_a)(\vec r_a \times \vec B)}{r^3_{a}}
- \frac{g_S-2}{4}
(\vec p_a\cdot \vec \sigma_a)\,(\vec p_a \cdot \vec B) \bigg\} 
\nonumber\\
& +\mu_B \alpha^2 \, \sum_{a \neq b}
\bigg\{-\frac{g_S - 1}{4}\,
\frac{(\vec r_{ab} \times \vec \sigma_a)(\vec r_a \times \vec B)}{r^3_{ab}}
- \frac{g_S}{4}\, 
\frac{(\vec r_{ab} \times \vec \sigma_b) (\vec r_a \times \vec B)}{r^3_{ab}}
+\frac{p_a^i}{2}
\bigg(\frac{\delta^{ij}}{r_{ab}} + \frac{r_{ab}^i\,r_{ab}^j}{r_{ab}^3}\bigg)\,
(\vec r_b \times \vec B)^j
\bigg\} \,, 
\nonumber\\
& \delta H_{\rm rel} = \mu_B\,\vec G_2 \cdot \vec B \,.
\end{align}
It is straightforward to identify 
the Cartesian tensor form of the relativistic correction $\vec G_2$,
\begin{align}
\label{G2}
G^i_2 =&\; \mu_B\,\alpha^2\,
\sum_a\,\bigg\{-
\frac{1}{2}\,(v_{2,a}^{i} + d_{2,a}\,\sigma^i_a)
+ \frac{Z\,(g_S - 1)}{4}
\bigg(\frac{2}{3}\,d_{3,a}\,\sigma_a^i - t_{3,a}^{ij}\,\sigma_a^j\bigg)
- \frac{g_S-2}{2}\,\bigg(\frac{d_{2,a}\,\sigma^i_a}{3} +
t_{4,a}^{ij}\,\sigma_a^j \bigg) \bigg\} \nonumber \\
&\; +\mu_B\,\alpha^2\,\sum_{a \neq b}\, \bigg\{ 
- \frac{g_S - 1}{4}\, \bigg(\frac{2}{3}\,d_{5,ab}\,\sigma_a^i - 
t_{5,ab}^{ij}\,\sigma_a^j + r_{5,ab}^{ij}\,\sigma_a^j \bigg)
- \frac{g_S}{4}\,\bigg(\frac{2}{3}\,d_{5,ab}\,\sigma_b^i - 
t_{5,ab}^{ij}\,\sigma_b^j + r_{5,ab}^{ij}\,\sigma_b^j \bigg)
\nonumber \\
&\; -\frac{1}{2}\,\left( v_{61,ab}^{i} - v_{62,ab}^{i} \right)
\bigg\}\,.
\end{align}
\end{widetext}
Indeed, the tensor components from the first four terms 
in Eq.~\eqref{G2} read as follows,
\begin{subequations}
\begin{align}
d_{2,a} =&\; \vec p_a^{\,2} \,,\\
v_{2,a}^{i} =&\; \vec p_a^{\,2}\,(\vec r_a \times \vec p_a)^i \,, \\
d_{3,a} =&\; \frac{1}{r_a} \,, \\
t_{3,a}^{ij} =&\; \frac{1}{r_a^3}\,\left(r_a^i r_a^j - 
\frac{1}{3}\,\delta^{ij}\,r_a^2\right) \,, \\
t_{4,a}^{ij} =& \; p_a^i\, p_a^j - \frac{1}{3}\,\delta^{ij}\,\vec p_a^{\,2} \,.
\end{align}
Furthermore, we have the following terms from the 
fifth corrections in Eq.~\eqref{G2},
\begin{align}
d_{5,ab} =& \; \frac{\vec r_{a} \cdot \vec r_{ab}}{r_{ab}^3} \,,\\
t_{5,ab}^{ij} =& \; \frac{1}{r_{ab}^3} 
\left( \frac{r_{a}^i r_{ab}^j + r_{a}^j r_{ab}^i}{2} - 
\frac{\vec r_{a} \cdot \vec r_{ab}}{3} \,\delta^{ij} \right)\,, \\
r_{5,ab}^{ij} =& \; \frac{1}{2 r_{ab}^3} \,
\left(r_{a}^i r_{b}^j - r_{a}^j r_{b}^i \right) \,.
\end{align}
Finally, the sixth term in Eq.~\eqref{G2}
yields a remaining vector structure,
\begin{align}
v_{61,ab}^{i} =& \; \frac{(\vec r_a \times \vec p_b)^i}{r_{ab}} \,, 
\\
v_{62,ab}^{i} =& \;
\frac{(\vec r_a \times \vec r_b)^i \, 
(\vec r_{ab} \cdot \vec p_b)	}{r_{ab}^{3}} \,.
\end{align}
\end{subequations}
We proceed to a final numerical evaluation of these corrections later.

%
% Self-energy correction
%
\subsection{Self-energy correction}
\label{se_corr}

We follow Ref.~\cite{Pa2004} and base the calculation of the 
low-energy part of the self energy 
proceeds on a nonrelativistic Hamiltonian in the presence 
of an electromagnetic field in the length gauge,
\begin{align}
H = H_0 + H_{M0} + H_\gamma - e\,\vec r_1 \cdot \vec E - e\,\vec r_2 \cdot \vec E\,,
\end{align}
where $H_0$ is the unperturbed Hamiltonian of the atom,
$H_\gamma$ is the Hamiltonian of the photon field,
$H_{M0}$ is the leading-order magnetic interaction
given in Eq.~\eqref{HM0},
and the two dipole interaction operators describe the 
interaction of the bound electrons with the 
quantized electromagnetic field.
The self energy has the form
\begin{align}
\delta E = &\; -\frac{2\,\alpha}{3\,\pi}\,
\int_0^{\epsilon} \dd \omega\,\omega^3\, 
\big \langle \phi \big |(\vec r_1 + \vec r_2) 
\nonumber \\
&\; 
\frac{1}{H_0 + H_M - E_0 + \omega}(\vec r_1 + \vec r_2) \big |\phi \big \rangle  \,.
\end{align}
It is understood that $\delta\,E $ is to be 
expanded in first order in the magnetic field $\vec B$.
Then, replacing the coordinates by electron momenta and using commutation
relations, it is easy to rederive 
Eq.~(38) of Ref.~\cite{Pa2004}, additionally assuming that the state
$\phi$ has definite $m_L$ and $m_S$ quantum numbers 
(projections of the orbital and spin angular momenta onto the 
quantization axis). We might just as well assume that 
the reference state has a defined value of the magnetic
quantum number $m_J$ of the total angular momentum,
\begin{align}
\label{ES}
\delta E =& \; 
-2\,\mu_B\,\frac{\alpha}{\pi}\,\biggl\{ \ii\,
\epsilon_{irs}\, \big \langle m_J\big | (\vec p_1 + \vec p_2)^r \nonumber
\\
& \; \times \ln|2(H_0-E)|\,(\vec p_1 + \vec p_2)^s\,\big |m_J  \big \rangle\,
\biggr\} \, B^i \,.
\end{align}
Adjusting the self energy correction to our convention, we obtain
the tensor structure
\begin{align}
\label{G3}
\delta E =& \; 
-2\,\mu_B\,\frac{\alpha}{\pi}\,
\big \langle m_J \big |\vec G_3 \cdot \vec B \big| m_J \big\rangle \,,
\end{align}
where $G_3^i\equiv v_7^i$ is given by the expression in curly brackets in 
Eq.~\eqref{ES}.  Thus, the self-energy correction has 
a simple vector structure, and contributes 
to the orbital momentum $L$ part.

%
% Spin and $\bm J$ Reduction
%
\section{Evaluation of the $\bm{g}$ Factor}
\label{GJ}

In first-order perturbation theory, one requires only the diagonal matrix
elements in the total angular momentum $\vec J$. Insofar as first-order theory
is concerned, one may replace $H_{M0} + \delta H_M$ by its restriction to the
$(2J+1)$-dimensional subspace spanned by the orthonormal vectors for
$m_J =-J,-J+1,\ldots, J$. Then, the magnetic Hamiltonian linear in the $\vec B$
field can be rewritten as follows, in terms  of the $g_J$-factor and $\vec J$,
\begin{equation}
H_{M0} + \delta H_M = 
\mu_B\,\sum_{\gamma = 0}^3 \vec G_\gamma \cdot \vec B = 
\mu_B\,g_J\,\vec J \cdot \vec B \,.
\end{equation}
In order to calculate the $g_J$ factor, it is helpful to
write it as an average over all magnetic projections. Using the 
shorthand notation $|m_J \rangle$ for the state with quantum numbers 
$|L,S,J,m_J\rangle$, an important relation is 
\begin{align}
\big \langle m_J \big | \vec G \cdot \vec B  \big | m_J \big \rangle = & \;
\left< m_J \left| \frac{\vec G \cdot \vec J \, \vec J \cdot \vec B}{\vec J^2} 
\right| m_J \right> 
\nonumber\\
=& \; m_J \,g_J \, B \,,
\nonumber\\[2ex]
g_J =& \; \frac{\big \langle m_J \big| \vec G \cdot \vec J \,\big 
| m_J \big\rangle}{J(J+1)} \,.
\end{align}
This relation holds for any $m_J$, and $\vec G$ may 
stand for any of the $\vec G_\gamma$ or for the 
sum $\vec G = \sum_{\gamma=0}^3 \vec G_\gamma$. Summing over $m_J$ and dividing 
by the number of states $2J+1$,
the $g_J$ factor can be determined as follows,
\begin{align}
\label{gJ}
g_J =& \; \frac{1}{J(J + 1) (2 J + 1)} 
\\[2ex]
& \; \times \sum_{i=0}\,\sum_{m_J = -J}^J 
\big \langle m_J \big | \vec G_i \cdot \vec J \, \big| m_J \big \rangle \,.
\nonumber
\end{align}
This expression involves a sum over the angular momentum projections
and is manifestly independent of $m_J$.

%
% Reduction of spin degrees and $\bm J$ 
%
\subsection{Reduction of spin degrees and $\bm J$ }
\label{reduction}

The $\vec G_\gamma$ with $\gamma = 0,1,2,3$
have been defined in Secs.~\ref{tensor},~\ref{rel_corr} and~\ref{se_corr}.
According to Eq.~\eqref{tensor2}, 
the terms  can be decomposed into diagonal $d$ terms, 
and symmetric as well as antisymmetric $t$ and $r$ terms.
Using Eq.~\eqref{gJ}, one can express the $g$-factor contribution
of a term of the form $\vec G \cdot \vec J$
in terms of the multiplicative factor $J(J + 1) (2 J + 1)$
a correction to the $g_J$ factor, and radial matrix elements.
For a contribution to $\vec G_\gamma \cdot \vec J$ of the form
$d\,\vec \sigma \cdot \vec J$,
one can deduce for $P$ states the formula
\begin{subequations}
\label{AJS}
\begin{align}
& \sum_{m_J =-J}^J 
\big \langle m_J  \big |\,d\,\vec \sigma \cdot \vec J \,\big| m_J \big \rangle = \\
&\qquad A_{JS}\,J(J+1)\,(2 J+1)\,
\sum_a\,\big \langle \psi^k \big| d \big| \psi^k \big\rangle \,,\nonumber \\
& A_{21}= A_{11} = \frac{1}{2}, \quad A_{10}=0 \,.
\end{align}
\end{subequations}
The Cartesian basis $| \psi^k \rangle$ of $P$ states is normalized to 
$\langle \psi^k | \psi^l \rangle  = \frac13\,\delta_{kl}$ 
[see Eq.~\eqref{proper_sym}]. 
The basis of the $| \psi^k \rangle$ states contains states without an explicit 
spin wave function, where the coordinate part is 
symmetrized or antisymmetrized,
according to Eq.~\eqref{proper_sym} below.
For a vector $\vec v$ coupled to $\vec J$ in Eq.~\eqref{gJ},
we use the following reduction scheme
\begin{subequations}
\label{BJS}
\begin{align}
\sum_{m_J=-J}^J & \;\langle m_J | \vec v \cdot \vec J | m_J \, \rangle = \\
&\;  B_{JS}\,J(J+1)\,(2 J+1)\,\ii\,\epsilon_{ijk} 
\langle \psi^i| v^j| \psi^k \rangle, \nonumber \\
B_{21}=& \; B_{11} = -\frac{1}{4}, \quad B_{10}=-\frac{1}{2} \,.
\end{align}
\end{subequations}
For a symmetric, traceless (quadrupole) tensor, we can project onto the 
Cartesian basis for $P$ states as follows:
\begin{subequations}
\label{CJS}
\begin{align}
& \sum_{m_J=-J}^J
\big \langle m_J \big |\, t^{ij}\,\sigma^i\,J^j \big | m_J \big \rangle = \\
& \qquad C_{JS}\,J(J+1)\,(2 J+1)\,
\sum_{a}\,\big \langle \psi^j \big | t^{jk} \big | \psi^k \big \rangle , 
\nonumber \\[2ex]
& C_{21}= -\frac{1}{10},\quad C_{11} = \frac{1}{2}, \quad C_{10}=0 \,.
\end{align}
\end{subequations}
Finally, for a antisymmetric tensor $r^{ij}$ coupled to $\sigma_a^i\,J^j$,
the total contribution to the $g_J$ factor vanishes for all states under
investigation here.

For excited helium $P$ states, the leading-order 
expression~\eqref{conv} evaluates to
\begin{equation}
g_J(n^3 P_{J=0,1,2}) = \frac32 \,, \qquad
g_J(n^1 P_1) = 1 \,,
\end{equation}
where $n$ is the principal quantum number of
the excited state, and the result for
$n^3 P_0$ is not of physical interest because the
magnetic projection for the state with $J=0$ always is $\mu = 0$.
The correction $\delta g_J$ to the
Land\'{e} $g$ factor can be expressed in terms of
$\delta g_L$, $\delta g_S$, and $g_x$
[see Eq.~\eqref{defdeltag}], and prefactors
$A_{JS}$, $B_{JS}$, and $C_{JS}$,
\begin{align}
\label{gJred}
\delta g_J =&\;\ A_{JS}\,\delta g_S - 2\,B_{JS}\,\delta g_L +
\frac{1}{3}\,C_{JS}\,\delta g_x \,,
\end{align}
where the $A_{JS}$, $B_{JS}$ and $C_{JS}$ coefficients are
given in Eqs.~\eqref{AJS},~\eqref{BJS} and~\eqref{CJS},
respectively.
It is probably useful to note 
that our scheme is easily generalized to other low orbital momentum states,
for example, $D$-states, which have submanifolds with $S=0$ and $S=1$, and $J = 1,2,3$.

%
% Table 1
%
\begin{table*}[!th]
\renewcommand{\arraystretch}{1.0}
\caption{Mean values of the tensor structures 
entering Eqs.~\eqref{AJS},~\eqref{BJS} and~\eqref{CJS}, 
for $2{}^1P$, $2{}^3P$ and $3{}^3P$ states.
In view of $A_{10} = C_{10} = 0$, only the phenomenologically
relevant results are indicated for singlet $P$ states.}
\label{table1}
\begin{ruledtabular}
\begin{tabular}{cw{2.16}w{2.16}w{2.16}}
Operator  & \cent{2\;{^1P}}& \cent{2\;{^3P}} & \cent{3\;{^3P}}\\
\hline \\
$\ii \,\epsilon_{ijk} \langle \psi^i | v^j_{0,a} |\psi^k\rangle$ & 
-2.0& -2.0& -2.0 \\
$\ii \,\epsilon_{ijk} \langle \delta \psi_{mp}^i | v^j_{0,a} |\delta \psi_{mp}^k\rangle$ & 
,-0.805\,549\,556(6) &-1.096\,171\,714(2) & -1.366\,172(4) \\
$\ii \,\epsilon_{ijk} \langle \delta \psi_{mp}^i |\delta \psi_{mp}^i\rangle$ & 
 0.402\,774\,778(8) & 0.548\,085\,857\,1(4) & 0.683\,086\,2(2) \\
$\ii\,\epsilon_{ijk} \langle \psi^i | v^j_{1,ab} |\psi^k\rangle$ & 
-0.131\,044\,018\,6(5) & 0.256\,875\,920\,7(3) & 0.069\,756\,861(4)\\
$\ii\,\epsilon_{ijk} \langle \psi^i | v^j_{1,a} |\delta \psi_{mp}^k\rangle$ &
-1.131\,383\,6(3) & -1.204\,232\,1(4) & -1.025\,36(6) \\
$\langle \psi^k | d_{2,a} | \psi^k \rangle$ & \multicolumn{1}{c}{$A_{10}=0$}
& 4.110\,292\,724\,2 & 4.116\,162\,168\,5 \\
$\ii\,\epsilon_{ijk} \langle \psi^i | v^j_{2,a} |\psi^k\rangle$ &
-0.483\,020\,291\,31 & -0.216\,764\,422\,66 & -0.259\,969\,064\,25 \\
$\langle \psi^k | d_{3,a} | \psi^k\rangle$ & \multicolumn{1}{c}{$A_{10}=0$}
& 2.109\,944\,701\,6 & 2.116\,045\,575\,2 \\
$\langle \psi^j | t^{jk}_{3,a} |\psi^k\rangle$ &  \multicolumn{1}{c}{$C_{10}=0$}
& 0.072\,236\,399(2) & 0.080\,111\,516\,9(4) \\
$\langle \psi^j | t^{jk}_{4,ab} |\psi^k \rangle$ & \multicolumn{1}{c}{$C_{10}=0$}
& 0.077\,581\,379(2) & 0.084\,694\,997\,4(3) \\
$\langle \psi^k | d_{5,a} |\psi^k\rangle$ & \multicolumn{1}{c}{$A_{10}=0$}
& 0.109\,596\,679\,06 & 0.115\,928\,981\,92 \\
$\langle \psi^j | t^{jk}_{5,ab} |\psi^k\rangle$ & \multicolumn{1}{c}{$C_{10}=0$}
& 0.066\,891\,418\,855(5) & 0.075\,528\,036\,275(1)\\
$\ii\,\epsilon_{ijk} \langle i | v^j_{61,ab} |\psi^k\rangle$ & 
-0.077\,009\,223\,65(3) & -0.024\,049\,685(6) & 0.030\,684\,751\,1(4)\\
$\ii\,\epsilon_{ijk} \langle \psi^i | v^j_{62,ab} |\psi^k\rangle$ & 
0.029\,866\,744\,798(1) & 0.008\,854\,469\,40(2) &-0.001\,716\,909\,955(6)\\
$\ii\,\epsilon_{ijk} \langle \psi^i | v^j_{7,ab} |\psi_{mp}^k\rangle$ & 
0.195\,754(2) & 0.264\,705(2) & 0.088\,415(4) \\
\end{tabular}
\end{ruledtabular}
\end{table*}

\begin{table}[t!]
\renewcommand{\arraystretch}{1.0}
\caption{$\delta g_J$ contributions to the singlet $2 \; {^1}P_1$ state
where $\delta g_S = \delta g_x = 0$. 
The fine-structure constant is $\alpha = 1/137.035\,999\,074(44)$, 
and the electron-$\alpha$ mass ratio is $m/M = 1.370\,933\,555\,78(55)$
(see Ref.~\cite{MoTaNe2008}).
Theoretical uncertainties come from our estimate of higher order 
effects in the order $\alpha^4$ (first parentheses), and finite-mass 
relativistic correction of order $\alpha^2 \lambda$ (second).}
\label{table2}
\begin{ruledtabular}
\begin{tabular}{lw{2.14}}
\rule[-2mm]{0mm}{6mm}
$2 \; {^1}P_1$ & \cent{\delta g_L \times 10^6 } \\
\hline 
Finite mass [Eq.~\eqref{G1mpol}] &   -8.968\,94  \\
Relativistic [Eq.~\eqref{Hrel}]              &   -7.853\,19  \\
Self-energy [Eq.~\eqref{G3}]                 &    0.024\,22  \\
Total             &  -16.798(9)(7) \\[2ex]
Theory: Ref.~\cite{AnSe1993}
                  &  -15.771       \\
Theory: Ref.~\cite{YaDr1994}        
                  &  -16.810\,165\ldots\footnote{There is 
no uncertainty estimate given in Ref.~\cite{YaDr1994}.}
\end{tabular}
\end{ruledtabular}
\end{table}

\begin{table*}[t!]
\renewcommand{\arraystretch}{1.0}
\caption{$\delta g_J$ contributions to triplet $P$ states.
Again, the theoretical uncertainty of the final 
theoretical prediction 
comes from the estimate of higher order
effects in the order $\alpha^4$, and finite-mass relativistic correction 
of order $\alpha^2 \, \lambda$ (first and second parentheses, respectively).}
\label{table3}
\begin{ruledtabular}
\begin{tabular}{lw{2.14}w{2.12}w{2.12}}
\rule[-2mm]{0mm}{6mm}
$2\;{^3P}_J$ & 
\cent{\delta g_L \times 10^6} & 
\cent{\delta g_S \times 10^6} & 
\cent{\delta g_x \times 10^6 } \\
\hline
Finite mass [Eq.~\eqref{G1mpol}]
                &   17.620\,32 &   0.0      &  0.0            \\
Relativistic [Eq.~\eqref{G2}]   
                &   -6.912\,92 & -80.429\,3 & -5.385\,487     \\
Self-energy [Eq.~\eqref{G3}]    
                &    0.032\,74 &   0.0      &  0.0            \\
Total           &   10.740(2)(2)  & -80.43(2)(4)  & -5.385\,5(13)(2)    \\[2ex]
Theory: Ref.~\cite{YaDr1994}         
 &   10.719\,291%
\ldots\footnote{There is no uncertainty estimate given in Ref.~\cite{YaDr1994}.} 
 &  -80.436\,904\ldots^a 
 &  -5.391\,808 \ldots^a \\
Theory: Ref.~\cite{LeHu1975}
                &   10.6(4)    & -80.46(1)  &  -3.5(1.5)      \\
Theory: Ref.~\cite{AnSe1993}
                &   8.838      & -80.401    &  -5.344         \\
Theory: Ref.~\cite{Pa2004}         
                &   10.752\,033\ldots  \\
Experiment: Ref.~\cite{LhFaBi1976}
                &   4.9(1.9)   &            &                 \\
Experiment: Ref.~\cite{LePiHu1970} 
                &   3.8(9.0)   &  -76.0(2.4)& 4.0(25.0)       \\
\hline
\hline
\rule[-2mm]{0mm}{6mm}
$3\;{^3P}_J$ & 
\cent{\delta g_L \times 10^6 } & 
\cent{\delta g_S \times 10^6 } & 
\cent{\delta g_x \times 10^6 }  \\
\hline 
finite mass [Eq.~\eqref{G1mpol}]
                &  4.788\,03 &   0.0       &  0.0           \\
relativistic [Eq.~\eqref{G2}]   
                & -3.029\,57 &  -75.083\,9 & -2.648\,665    \\
self energy [Eq.~\eqref{G3}]    
                &  0.010\,94 &   0.0       &  0.0           \\
Total           &  1.769(9)(2)  & -75.08(2)(4)   & -2.648\,7(7)(3)  \\[2ex]
theory: Ref.~\cite{YaDr1994}
                &  1.772\,223\ldots^a  & -75.096\,557\ldots^a & -2.650\,192\ldots^a\\
theory: Ref.~\cite{KrPi1978}
                &  -0.17(2.8)& -75.13(3.27)& -2.75(10.02) 
\end{tabular}
\end{ruledtabular}
\end{table*}
%%%%%%%%%%%%%%%%%%%%%%%%%%%%%%%%%%%%%%%%%%%%%%%%%%%%%%%%%%%%%%%%
%%%%%%%%%%%%%%%%%%%%%%%%%%%%%%%%%%%%%%%%%%%%%%%%%%%%%%%%%%%%%%%
%%%%%%%%%%%%%%%%%%%%%%%%%%%%%%%%%%%%%%%%%%%%%%%%%%%%%%%%%%%%%%%

%
% Numerical Evaluation
%
\section{Numerical Evaluation}
\label{num}

The nonrelativistic wave function of the $P$ state $\psi$ and
its energy $E_0$ are determined for a Schr{\"o}dinger equation 
with a nonrelativistic Hamiltonian Eq.~\eqref{H0}
\begin{align}
H_0 \, \psi =&\; E_0 \, \psi,
\end{align}
based on the Rayleigh-Ritz variational principle. We use a basis set of explicitly 
exponentially correlated functions (following 
Refs.~\cite{Ko2000,PuJeMo2011})
\begin{align}
\label{proper_sym}
\psi^k =& \; \sum_{m=1}^{N} w_m \; \big[r^k_1  e^{-a_k r_1 - b_k r_2 - c_k
r_{12} } \mp (r_1 \leftrightarrow r_2) \big],
\end{align}
which for the singlet (triplet) states is symmetric (antisymmetric)
under an exchange of spatial coordinates, as required by the 
Pauli exclusion principle. We reemphasize that the 
superscript $k$ denotes the Cartesian coordinate, that is,
the wave function with an orientation along the $x$ axis 
would be denoted as $\psi^x$ and involve the 
$x$ coordinates $r_1^x$ and (in the exchange term) $r_2^x$. 
The parameters $(a,b,c)$ 
for the $i$th function are randomly generated from an optimized box 
$(A_1,A_2) \times (B_1,B_2)\times (C_1,C_2)$ 
under the additional constraints $a_k + b_k > \varepsilon$ as well as
$b_k + c_k > \varepsilon$ and $c_k + a_k > \varepsilon$, where 
$\varepsilon = \sqrt{2 \, ( E_0^{+} - E_0)}$ with $ E_0^{+}$ being the lowest 
singlet (triplet) $P$ state energy for He$^{+}$. 

In order to obtain a more accurate 
representation of the wave function, we use two boxes that model the short-range 
and medium-range asymptotics of the helium wave functions. In this basis, 
the matrix element of the nonrelativistic Hamiltonian $H_0$ can be 
represented as a linear combination of the integrals
\begin{align}
& \Gamma(a,b,c,n_1,n_2,n_{12}) \\[2ex]
& = \int \dd^3 r_1 \; \dd^3 r_2 \;
r_1^{n_1 - 1} r_2^{n_2 - 1} r_{12}^{n_{12}- 1} \,
\ee^{-a r_1 - b r_2- cr_{12}} \,,
\nonumber
\label{gamma}
\end{align}
with non-negative $n_1$, $n_2$, and $n_{12}$. Methods
for their computation are well known \cite{SaRoKo1967}. The linear coefficients 
$d_m$ in Eq. (\ref{proper_sym}) are obtained from a solution of a generalized 
eigenvalue problem. The numerical accuracy of the results is estimated 
from the apparent numerical convergence of the matrix elements as the 
size of the basis is increased. For the calculation of $\psi^k$, we use
an expansion with a moderate number up to 
$2\,N = 900$ basis functions (we use a prefactor $2$ 
in order to clarify the distribution of the basis functions onto two 
variational boxes, as described in Ref.~\cite{PuJeMo2011}).
The numerical accuracy of the following nonrelativistic 
reference values,
\begin{subequations}
\label{entries}
\begin{align}
E_0(2\,^1 P) =&\; -2.123\,843\,086\,498\,101\,3(3) \,, \\
E_0(2\,^3 P) =&\; -2.133\,164\,190\,779\,283\,1(2) \,, \\
E_0(3\,^3 P) =&\; -2.058\,081\,084\,274\,275(1) \,,
\end{align}
\end{subequations}
is fully sufficient for our calculations.
All entries in Eq.~\eqref{entries} are consistent with 
the values given in Table~III of Ref.~\cite{Dr1993ryd}.
For $2\,^3 P$, our result also is in agreement with the entry 
in Eq.~(20) of Ref.~\cite{Pa2002}.
 
Using the wave functions $\psi^k$, we are able to obtain all necessary mean
values defined by the tensor components $d$, $v^i$ and $t^{ij}$. 
In order to perform the calculation, the set of integrals of the form 
given in Eq.~\eqref{gamma} needs to be extended by
additional classes including one or two of the indices $n_1$, $n_2$, and
$n_{12}$ being equal to $-1$. The analytic formulas for the master integrals
and related recurrence schemes are well known, as given in Ref.~\cite{Ko2002jpb}.
Numerical values for the states with definite orbital momentum and
spin-angular symmetries $2{}^1P$, $2{}^3P$ and $3{}^3P$ are presented in 
Table~\ref{table1}. Because the coefficients $A_{10}$ and
$C_{10}$ vanish, we do not provide values for the $d$ and $t^{ij}$ 
elements for 
singlet $P$ states. Values without an error estimation are cut to eleven 
digits; all of these are believed to be numerically significant.

In order to determine the finite mass effect of order $O(\lambda^2)$, 
we first calculate the mass polarization correction to the wave function,
scaling the $\lambda$ parameter out of the perturbation,
\begin{align}
\label{mpol}
\lambda\,\big |\delta \psi_{mp}^k \big \rangle = 
-\lambda\,\frac{1}{(E_0 - H_0)^{'}} \, 
\frac{\vec p_1 \cdot \vec p_2}{\mu}
\big | \psi^k \big \rangle \,.
\end{align}
The expression $| \delta \psi_{mp}^k \rangle$ 
is relevant for the entries in the third column of Table~\ref{table1}.
The operator $\vec p_1 \cdot \vec p_2$ in Eq.~\eqref{Hmp} changes
neither the orbital angular momentum nor 
the spin symmetry when acting on $\psi^k$ in
Eq~\eqref{mpol}. Thus, it can be expressed using a basis consisting only
of the $\psi^k$ defined in 
Eq.~\eqref{proper_sym}. Variational parameters for $\delta \psi_{mp}$ are
generated in analogy to those for the wave function~$\psi$, but the size  of the
basis is chosen to be larger ($2 N_{mp} = 3 N$). 
With these results in hand, it
is straightforward to calculate the mass polarization correction for a
given operator. The only effect is the second-order correction to the $v^i_{1,ab}$
in Eq.~\eqref{G1}. Together with part of the third order correction to $v^i_{0,a}$, we obtain 
\begin{align}
\label{G1mpol}
\langle \psi^i | G_1^j |\psi^k\rangle = 
-\frac{m}{M}\,\sum_{a \neq b}\,&\;
\big[\langle\, \psi^i | v_{1,ab}^{j} |\psi^k\rangle  \\ 
&\; + 2\,\lambda \,
\langle \psi^i | v_{1,ab}^{j} | \,
\delta\,\psi^k_{mp}\rangle \big]\,\nonumber \\
+\,\lambda^2\,\sum_{a}\,&\;
\langle \delta \psi_{mp}^i | v_{0,a}^{j} |\delta \psi_{mp}^k\rangle, 
\nonumber
\end{align}
where the formula is expressed in compact form by writing the 
coefficient of the first term as $m/M$, not $\mu/M$.
The other terms in the third order perturbation of $v^i_{0,a}$ result in the
shifted $g_L$ coefficient in the leading term in Eq.~\eqref{HM}
%k
\begin{align}
\label{gLshift}
g_L =&\; 1 - \frac{m}{M} - 
\lambda^2\,\langle \delta \psi_{mp}^i | \delta \psi_{mp}^i\rangle \nonumber \\
=&\; 0.999\,862\,916\,942\,649(5)(55) \,,
\end{align}
where the first uncertainty estimate refers to the numerical uncertainty 
of the particular contribution (finite-mass correction to $g_L$), and
the second uncertainty comes from the CODATA electron-$\alpha$ mass ratio 
$m/M = 1.370\,933\,555\,78(55)$.

The most numerically intensive part of the $g$ factor calculations in Eq.~\eqref{gJ} 
is the vector component of the self-energy correction~\eqref{G3}, which 
can be reduced~\cite{Pa2004} to the expression 
\begin{align}
\label{logHE}
\ii \epsilon_{ijk} \big \langle \psi^i \big |v_7^j&\;\big| \psi^k \big\rangle = 
(\delta^{ir}\delta^{ks}-\delta^{is}\delta^{kr})\,
\big \langle \psi^i \big | (\vec p_1 + \vec p_2)^r
\nonumber \\
& \; \times \ln|2(H_0-E)|\,(\vec p_1 + \vec p_2)^s\,\big |\psi^k \big \rangle \,,
\end{align}
which differs from the ordinary Bethe logarithm in the 
absence of a linear term $H_0-E$ multiplying the logarithm.
Matrix elements involving the logarithm of the Hamiltonian
necessitate the use of the methods usually employed 
for Bethe logarithm calculations 
for excited states in helium, where due to 
conceivable numerical challenges, for a long time
asymptotic formulas~\cite{Sc1961helium,DrSw1990,GoDr1992,DrMa1998} in
$1/n$ and $1/Z$ were the preferred method of 
calculation. Direct and accurate calculations
of logarithmic sums over the helium spectrum 
have become possible only quite 
recently~\cite{DrGo1999,Ko1999,Ko2004}. 
Here, we closely follow to the integral representation 
of the Bethe logarithm~\cite{PaKo2004}, which for 
the expression in Eq.~\eqref{logHE} has a particular compact form,
\begin{subequations}
\begin{align}
\ii \,\epsilon_{ijk} \big \langle \psi^i \big |v_7^j&\;\big| \psi^k \big\rangle = 
\,\int_{0}^1\,\dd t\,\frac{f(t)}{t^3} \,, 
\\[0.77ex]
\label{ft}
f(t) = (\delta^{ir}\delta^{ks}-&\;
\delta^{is}\delta^{kr})\big \langle \psi^i \big | (\vec p_1 + \vec p_2)^r  
\\ 
&\; \frac{1}{H_0-E_0+\omega}\,
(\vec p_1 + \vec p_2)^s\,\big |\psi^k \big \rangle \,,
\nonumber\\[0.77ex]
t =&\;\frac{1}{\sqrt{1  + 2\,\omega}}.
\end{align}
\end{subequations}
We perform an integration over 100 equally spaced and 
optimized $t$ points, following ideas outlined in Ref.~\cite{PaKo2004}. 
The well-defined limit of the integrand, 
$\lim_{t \rightarrow 0} f(t)/t^3 = 0$,
facilitates the numerical evaluation.

%%%%%%%%%%%%%%%%%%%%%%%%%%%%%%%%%%%%%%%%%%%%%%%%%%%%%%%%%%%%%%%%
%%%%%%%%%%%%%%%%%%%%%%%%%%%%%%%%%%%%%%%%%%%%%%%%%%%%%%%%%%%%%%%
%%%%%%%%%%%%%%%%%%%%%%%%%%%%%%%%%%%%%%%%%%%%%%%%%%%%%%%%%%%%%%%
%
% Results
%
\section{Results}
\label{res}

The numerical data for the individual operators allows us to obtain theoretical
analysis of the $L$, $S$ and the $x$ 
part of the Land\'{e} $g$ factor,
as defined in Refs.~\cite{LePiHu1970,YaDr1994}
and discussed above. We express
our results in terms of $\delta g_J$,
which is obtained as the difference of the total 
prediction and the leading term [see Eq.~\eqref{defdeltag}].
For the numerical evaluation, we use Eq.~\eqref{gJ}.
We express the correction $\delta g_J$ to the 
Land\'{e} $g$ factor in terms of 
$\delta g_L$, $\delta g_S$, and $\delta g_x$, and prefactors
$A_{JS}$, $B_{JS}$, and $C_{JS}$, 
as given in Eqs.~\eqref{AJS},~\eqref{BJS} and~\eqref{CJS}, and~\eqref{gJred}.

We keep the conventions of Refs.~\cite{LePiHu1970,YaDr1994}
and compare our results to the experimental and
theoretical literature. In Tables~\ref{table2} 
and~\ref{table3}, we provide data split into a finite-mass
part related to Eq.~\eqref{G1} including the second order mass polarization
correction Eq.~\eqref{G1mpol}, the relativistic correction Eq.~\eqref{G2}, and
the self-energy correction 
given in Eq.~\eqref{G3}. The first conceivable
source of uncertainty for these
contributions is purely numerical, due to the finite 
numerical accuracy of the components in
Table~\ref{table1}. However, in most cases, the 
numerical uncertainty is negligible as
compared to the omitted higher-order effects.

We are able to report that our results
confirm the numerical data reported 
previously in Ref.~\cite{YaDr1994} at the
level of relativistic operators without finite-mass 
corrections. Here, we attempt to go beyond the 
leading relativistic effects. 
The self-energy correction to the $P$ state $g$ factor 
consists of two parts, one of which involves a Bethe-logarithm 
type term (logarithmic sum over virtual excited states) and 
is ultraviolet finite, in contrast to the Bethe logarithm
contribution to the $S$ state Lamb shift, which is 
known to be ultraviolet divergent~\cite{Be1947,Pa1993}.
The second contribution due to the self-energy is a 
high-energy contribution, which is manifest
in the anomalous magnetic effects of the free electron,
which are included into the 
relativistic Zeeman Hamiltonian 
given in Eq.~\eqref{Hrel}.
These contributions are infrared finite. 
We obtain very good agreement with a
numerical result reported in Ref.~\cite{Pa2004} 
for the self energy correction $\delta g_L$ for $2^3P$.
Combining relativistic and radiative effects,
we should mention the presence of an 
additional third term with a prefactor $g_S-2$ in Eq.~\eqref{Hrel}.
In Ref.~\cite{YaDr1994}, this term had not been taken into account,
apparently, but its numerical magnitude does not 
shift the final result significantly.
Our self-energy correction, which we add
to the relativistic result of
Ref.~\cite{YaDr1994}, includes the relativistic anomalous magnetic moment
effects of the electron and the spin-dependence of the 
Bethe logarithm term.
We also include $O(\lambda^2)$ corrections to the leading order 
resulting in the finite mass correction in Eq.~\eqref{G1mpol}. 
Such correction gives $-0.8 \times 10^{-8}$, $-1.0 \times 10^{-8}$ 
and $-1.3 \times 10^{-8}$ to the $\delta g_L$ of $2^1P$, 
$2^3P$ and $3^3P$ respectively, having the order of the self energy correction.

There are two significant sources of theoretical uncertainty for the final
results, which are given by finite-mass corrections to the relativistic effects
and by higher-order QED contributions.  In the results reported in
Ref.~\cite{YaDr1994}, the mass scaling and mass polarization corrections to the
relativistic effects have been included; these results are of relative order
$O(\alpha^2 \, \lambda)$ with respect to the leading $g$ factor term. However,
there are additional finite mass relativistic effects of the same order, which
can be deduced from Eq.~(40) in Ref.~\cite{Pa2008}, and which should 
be included in a systematic treatment. Here, 
we do not perform a complete
calculation of the terms of order $O(\alpha^2 \, \lambda)$,
and so we do not include the relativistic reduced-mass correction at all in our
final results.  We use some partial results we have obtained in the order 
$\calO(\alpha^2 \, \lambda)$ in order to estimate the size of the relativistic-recoil
correction. 

These include the scaling and the mass polarization corrections for
the $L$ part in $2^1P_1$ ($4.4 \times 10^{-9}$), as well as for the $x$ part
($1.6\times 10^{-9}$) and the $S$ part ($2.5\times 10^{-8}$).  For 
$2^3P_J$, we have $1.1\times 10^{-10}$ for the $x$ part, 
$9.2\times 10^{-10}$ for the $L$ part,
and $2.5\times 10^{-8}$ for the $S$ part.  Finally, for the
$x$ part in $3^3P_J$, we have a result of $1.7\times 10^{-10}$.  These 
results guide our estimates of the theoretical uncertainty indicated in
Tables~\ref{table2} and~\ref{table3}, where we multiply the 
partial results with a conservative 
weight factor of~$1.5$.  For the
higher-order QED contributions, we are not even able to present approximate
formulas based on, for example, 
hydrogenic contributions, because the theory of the Zeeman
effect has not been developed until now to this order. 
Therefore, we use the combined value from
the relativistic correction multiplied by a factor $\alpha^2$, and 
the leading QED correction multiplied by a factor $\alpha$,
to obtain a conservative estimate of the
uncertainty due to the uncalculated 
higher-order effects. We employ an additional conservative enlargement
factor of $5$ in order to estimate the size
of the effects of order~$\alpha^3$. 

As evident from Tables~\ref{table2} and~\ref{table3}, 
agreement of theory and experiment is rather satisfactory 
for a number of contributions, with the 
exception of a $2\,\sigma$ discrepancy for $\delta g_S$
(comparing our result to that of Ref.~\cite{LePiHu1970})
for the $2 {}^3P$ state and a 
$3\,\sigma$ discrepancy for $\delta g_L$
(comparing our result to that of Ref.~\cite{LhFaBi1976}).
It would be very interesting to remeasure the 
effect and clarify the status of the experimental results.

%
%
% Conclusions
%
\section{Conclusions}
\label{conclu}

In the calculation of the $g_J$ factor of 
excited states of helium, all aspects of atomic 
physics play a role: electron correlation, relativity,
and QED radiative corrections. Furthermore, these
effects are all intertwined, that is, there are 
QED radiative corrections multiplying the 
relativistic effects, contributing in higher order.
We here carry out a theoretical analysis of the 
$P$ state $g$ factor of singlet and triplet helium 
states, with a special emphasis on 
relativistic, radiative, and finite-mass 
corrections. The effects are calculated
through order $\alpha^2$ (for the relativistic effects),
and we also include radiative effects
of order $\alpha^3$. Furthermore, finite-mass 
corrections of order $\lambda$ and $\lambda^2$ are included,
while available partial results for the 
effects of relative order $\alpha^2 \, \lambda$ 
are used in order to estimate the theoretical uncertainty
in this order.
Results are summarized in Tables~\ref{table2} and~\ref{table3};
a $2\,\sigma$ discrepancy for $\delta g_S$
in comparison to Ref.~\cite{LePiHu1970} and a
by now-famous $3\,\sigma$ discrepancy for $\delta g_L$
in comparison to Ref.~\cite{LhFaBi1976} highlight the need 
for additional experimental evidence before definitive 
conclusions can be drawn.

With regard to the QED self-energy correction
to the $g$ factor, one encounters a peculiar situation
for $P$ states: Namely, both the high-energy part as well 
as the low-energy part are separately finite,
while the low-energy part is given by a Bethe-logarithm 
term, and the high-energy part is given by the 
anomalous magnetic moment. Both terms are finite,
and it is easy to overlook one of the contributions.
For atomic hydrogen, 
this has been verified both numerically~\cite{YeJe2010} 
and analytically~\cite{Je2010g}.
For helium singlet versus triplet states,
we find that the low-energy part of the 
self-energy correction to the bound-state
$g$ factor is spin-dependent.

Our calculations are performed in an angular-momentum 
coupling scheme which allows us to separate the 
internal degrees of freedom of the atom from the 
interaction with the external magnetic field.
The angular momentum algebra can become rather involved
for helium $P$ states.
We fully confirm the relativistic treatment of Ref.~\cite{YaDr1994}
using our mixed approach.
In a more general context, one may recall that the 
contribution of the anomalous magnetic moment 
to the $P$ state lifetime in few-electron 
systems has recently given rise to interesting 
experimental-theoretical discrepancies~\cite{LaEtAl2005}
which remain to be fully resolved.

%
% Acknowledgments
%
\section*{Acknowledgments}

This work was supported by the National Science Foundation
(grant PHY-1068547) and by a precision measurement grant
from the National Institute of Standards and Technology.
The authors acknowledge helpful conversations with K.~Pachucki.
Support from Pozna{\'n} Networking and Supercomputing Center also is 
gratefully acknowledged.

\appendix 

%
% Nonrelativistic Treatment
%
\section{Nonrelativistic Treatment}
\label{appa}

The nonrelativistic Hamiltonian of an $n$-electron atom is given as 
(in atomic units)
\begin{equation}
H = \frac{\vec p_N^2}{2M} + \sum_a
\left( \frac{\vec p_a^2}{2 m} - \frac{Z}{r_a} \right) + 
\sum_{a > b} \frac{1}{r_{ab}} \,,
\end{equation}
where we keep the electron mass $m$ and the nucleus mass $M$
in symbolic form.
In the center-of-mass system, we have 
$\vec p_N = - \sum_a \vec p_a$, and therefore
\begin{equation}
\label{A2}
H = \sum_a \left( \frac{\vec p_a^2}{2\mu} - \frac{Z}{r_a} \right) 
+ \sum_{a > b} \left( \frac{1}{r_{ab}} + 
\frac{\vec p_a \cdot \vec p_b}{M} \right) \,,
\end{equation}
where the latter term corresponds to the mass polarization.
The reduced mass $\mu$ is given as
\begin{equation}
\frac{1}{\mu} = \frac{1}{m} + \frac{1}{M} \,.
\end{equation}
If we define the ratio $\lambda = -\mu/M$ as in 
Eq.~\eqref{deflambda}, then an important identity is 
\begin{equation}
1 + \lambda = 1 - \frac{\mu}{M}
% = 1 - \frac{m}{m+M} 
= \frac{\mu}{m} \,.
\end{equation}

\end{document}